# DEEP BV: A FULLY AUTOMATED SYSTEM FOR BRAIN VENTRICLE LOCALIZATION AND SEGMENTATION IN 3D ULTRASOUND IMAGES OF EMBRYONIC MICE

*Ziming Qiu[1], Jack Langerman[2], Nitin Nair[1], Orlando Aristizabal[3,4], Jonathan Mamou[3], Daniel H. Turnbull[4], Jeffrey Ketterling[3], Yao Wang[1]*

[1] Electrical and Computer Engineering, Tandon School of Engineering, New York University, Brooklyn, USA
[2] Computer Science, Tandon School of Engineering, New York University, Brooklyn, USA
[3] F. L. Lizzi Center for Biomedical Engineering, Riverside Research, New York, USA
[4] Skirball Institute of Biomolecular Medicine, New York University School of Medicine, New York, USA

{zq415, jackmlangerman, nn1174, yw523}@nyu.edu, {orlando.aristizabal, daniel.turnbull}@med.nyu.edu
{JMamou, Jketterling}@riversideresearch.org

**Abstract** — Volumetric analysis of brain ventricle (BV) structure is a key tool in the study of central nervous system development in embryonic mice. High-frequency ultrasound (HFU) is the only non-invasive, real-time modality available for rapid volumetric imaging of embryos in utero. However, manual segmentation of the BV from HFU volumes is tedious, time-consuming, and requires specialized expertise. In this paper, we propose a novel deep learning based BV segmentation system for whole-body HFU images of mouse embryos. Our fully automated system consists of two modules: localization and segmentation. It first applies a volumetric convolutional neural network on a 3D sliding window over the entire volume to identify a 3D bounding box containing the entire BV. It then employs a fully convolutional network to segment the detected bounding box into BV and background. The system achieves a Dice Similarity Coefficient (DSC) of 0.8956 for BV segmentation on an unseen 111 HFU volume test set surpassing the previous state-of-the-art method (DSC of 0.7119) by a margin of 25%.

## I. INTRODUCTION

The mouse's ubiquity as an animal model in the study of mammalian development is due to the high degree of genome homology between mice and humans. One of the key methods for tracking genetic mutations is to observe how these mutations manifest themselves during embryonic development as variation in the shape of the brain ventricle (BV) in 3D views [1]. High-throughput, high-frequency ultrasound (HFU) has proved to be a promising imaging modality for phenotyping mouse embryos due to its fast 3D data-acquisition capability and the wide availability of commercial and research ultrasound scanners [2]. However, manual segmentation is very time-consuming with segmentation of a single HFU scan requiring around fifteen minutes of trained expert work. Therefore, it is essential to develop fully automatic segmentation algorithms [3].

Some typical HFU images of mouse embryos are shown in Fig. 1 and Fig. 5. As can be seen, there is great variation in BV location and shape as well as body posture, depending on probe positioning and developmental stage of the embryo. In addition, the boundary between the BV and the surrounding regions can be blurry due to motion (Fig. 5–red arrows). Furthermore, specular reflection or shadowing from overlying tissues may lead to loss of boundary contrast (Fig. 1–white arrow).

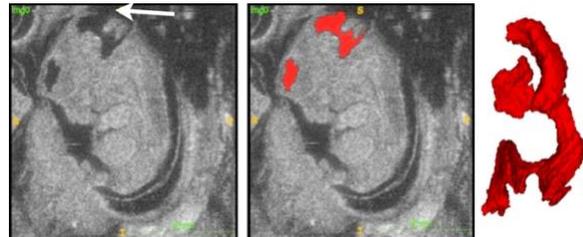

**Figure 1**: A slice of a sample 3D HFU volume (note the missing head boundary indicated by the white arrow) (left), Deep BV predicted segmentation (center), and 3D view of the segmented BV (right) –Dice Similarity Coefficient for this sample is 0.914.

A previous work [4] introduced a dataset of 36 HFU volumes with manual BV segmentations. In this paper, 370 additional whole-body image volumes with manual BV segmentations are used. This expanded dataset is divided into 259 samples for training and 111 for testing.

Nested Graph Cut (NGC) [5] was first proposed to segment the brain ventricle from an image of a mouse embryo head (manually cropped from the whole-body image). NGC makes use of the nested structure of the head image and is able to overcome the missing head boundary problem. This problem is caused by a loss of ultrasound signal due to either specular reflection or shadowing from overlaying tissues. Subsequent work [4] focused on BV segmentation in whole-body images. The method described in [4] first detects the embryo surface and then applies NGC to the embryo region to segment the BV. Although this framework performed well on the 36 HFU image volumes it was tuned on [4], subsequent testing on the unseen 111 image test set gave a DSC of 0.7119. We suspect that this is because the framework was optimized for the original 36 images and does not generalize well to new data. A comparison of mean DSC on the unseen test set of the deep learning based method proposed in this work and the NGC based framework [4] is presented in Table 2.

With the availability of powerful modern computation resources and large scale labeled data, deep learning has shown enormous success in various computer vision tasks, including biomedical image analysis [6][7]. Milletari et al. [8] successfully applied Hough voting-based convolutional neural networks (CNNs) to localize and segment the midbrain in MRI and Ultrasound images by voting based on results from patch level classifiers. Even though this method uses Hough voting to incorporate a shape prior implicitly, the patch-wise training strategy ignores the relationships between neighboring patches while training the CNN classifiers. Long et al. [9] first develop a fully convolutional network (FCN) by removing all the fully connected layers from traditional CNN based classifiers and introducing learnable up-sampling filters via the transpose convolutional layer. This enables end-to-end and pixel-to-pixel training for natural image semantic segmentation. This model greatly improves image semantic segmentation and has become pervasive serving as an inspiration for many related works in a variety of applications. Ronneberger et al. [10] modify the FCN model to introduce symmetric skip connections, which aim to combine high-resolution features (low level but good for localization) and low-resolution features (high level and good for semantic meaning). This U-net style architecture has been successfully used in a wide variety of biomedical applications and has achieved promising results.

Unlike natural images, many of the principal modalities in medical imaging are inherently three dimensional (CT, MR, Ultrasound, etc.), and as such, volumetric image segmentation is a key task in biomedical image processing. Milletari et al. [11] further adapt U-net to V-net for volumetric medical image segmentation. They also introduce a Dice-based loss function which, intuitively, can focus on volumetric overlap rather than treating each voxel as a separate binary classification problem. Directly applying V-net to our BV segmentation task is ineffective because there is an extreme imbalance between background and foreground (i.e. the BV makes up only 0.335% of the whole volume on average). Additionally, we identify four primary challenges for BV segmentation from whole-body images: (1) variation in body posture and orientation; (2) differing BV shapes and locations; (3) presence of severe missing head boundaries and motion artifacts (see Fig. 5), and (4) variation in image sizes from 150×161×81 to 300×281×362.

In this paper, we propose a fully automatic and robust deep learning based BV segmentation framework, hereafter referred to as Deep BV (Fig. 2). The system has two primary components: Localization and Segmentation. First, a 128×128×128 3D sliding window is used to scan the input image with a volumetric CNN as a classifier to obtain a bounding box. This serves as a hard attention mechanism. Then, a 3D fully convolutional network is used to classify each voxel within the detected bounding box as BV or background.

The main contribution of this work is the successful development of a fully automatic framework using deep learning that achieves state-of-the-art BV segmentation results. Compared to the previous state-of-the-art, a traditional graphical model based method [4], our proposed deep learning based framework is much more robust to variation in embryo body orientation, BV shape and BV location. It can obtain satisfactory results even when the image has highly inconsistent intensity distributions and severe missing boundaries.

To the best knowledge of the authors, this is the first Deep Learning based pipeline for segmentation in HFU scans of mouse embryos. In particular, the novelty of our pipeline lies in the use of a sliding window localization network to act as hard attention mechanism prior to segmentation, and the extension of V-Net with the addition of Large Kernels[12], Residual Connections[13], a full resolution stream [14], and cross constrained filter structures.

## II. METHODS

As illustrated in Fig. 2, our fully automated framework consists of two modules: BV localization and segmentation. The structure of each module will be described in detail in this section.

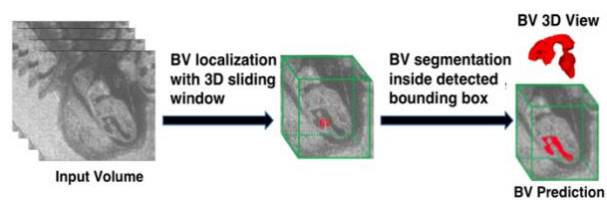

**Figure 2**: The proposed Two-part Deep BV pipeline: (i) A region containing the BV is located by applying a binary classifier to a sliding window over the full input volume. (ii) Then the segmentation model further identifies BV voxels within the detected region.

### A. LOCALIZATION

BVs have large variation across the 259 training images in location, orientation, and size ranging from 28 to 129 voxels per side. We take advantage of the observation that each scan only contains a single BV by using a 3D sliding window of size 128×128×128, which is large enough to contain even the largest BV, to scan the whole input volume for localizing the BV in the whole-body image. When the input image size along any side is smaller than 128, we zero pad the image to a size of 128 in that dimension. For computational efficiency, this process is performed at half resolution (i.e., a 64×64×64 sliding window is used to scan images after down sampling by a factor of 2). We adopt a 10-layer VGG-style network architecture [15] as the classifier (illustrated in Fig. 3). In the network, each of the convolutional and linear layers are followed by Rectified Linear Unit (ReLU) nonlinearities [16] and Batch Normalization [17]. To ensure robustness of the network and protect against overfitting, dropout layers [18] are inserted following each max pooling layer with dropout rate of 0.15 and after the first linear layer with dropout rate of 0.4. To train the CNN classifier, we define all windows which contain less than 80% of the BV to be negative examples, and all windows which contain greater than 99% of the BV to be positive examples. The remaining windows are considered ambiguous and are not used in training. In the down-sampled image, stride 2 is used for extracting positive training examples and stride 3 is used for negative examples.

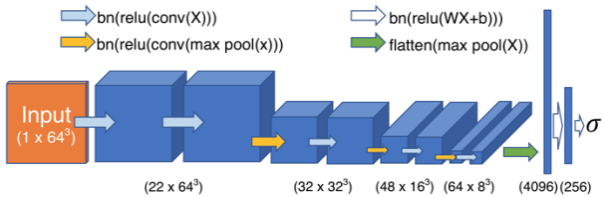

**Figure 3**: Pictorial representation of the localization network.

B. SEGMENTATION

When designing the segmentation architecture, we sought to address two primary design challenges:

1) Over-parameterization: Given the relatively small size of datasets in medical imaging, and the volumetric nature of many medical imaging modalities, the problem described by many including [19] of overly rapid parameter growth, is especially acute. This results from the fact that the number of parameters is cubic in the filter size rather than quadratic as is the case with two dimensional inputs such as natural images.

2) Sufficient field of view to account for global structure: The network should be deep enough (and kernels large enough [12]) so that each activation in the later stages of the network has the potential to incorporate information from the whole input volume. This is crucial because it is quite difficult (even for experts) to determine whether a dark region belongs to the BV without examining the surrounding visual context and considering global structure.

To address these considerations, a U/V-Net [10][11] style encoder-decoder architecture is employed, which allows for the incorporation of high level (deeper) features which contain global structural information and low level (shallower) features for pixel-level segmentation. However, there are three key differences between the segmentation architecture for Deep BV (see Fig. 4) and that of a traditional U/V-Net.

1) In V-Net, the input is fed through four down sampling blocks followed immediately by four up sampling blocks [11]. However, we opt to add seven additional Low Resolution Processing (LRP) layers at the lowest resolution. Expanding the depth of the network at this resolution results in a much more rapid gain in effective receptive field size with each additional layer thus facilitating the extraction of features which contain global structural information.

2) Spatially separated / cross constrained filter structures are used in LRP layers to combat excessive parameter growth as described in [19] resulting in a deeper structure with many fewer parameters.

3) A full resolution processing stream [13][14] is added with three full resolution convolutional layers at the beginning of the network, the outputs of which are concatenated with the result of the deep stream of the network as input to a full resolution border refinement and post processing stage.

All convolutional layers use 3D kernels of size $7 \times 7 \times 7$. All cross constrained filters are formed using sums of three orthogonal 1D filters of length 7. Using cross constrained convolution in the LRP layers reduces the parameter count from being cubic in the kernel size to linear in the kernel size. For the Deep BV segmentation architecture (Fig. 4), this results in a reduction from greater than 15M parameters to less than 2M. As demonstrated in the experimental section this constrained filtering structure leads to very accurate BV segmentation while substantially reducing the computation needed both for training and inference.

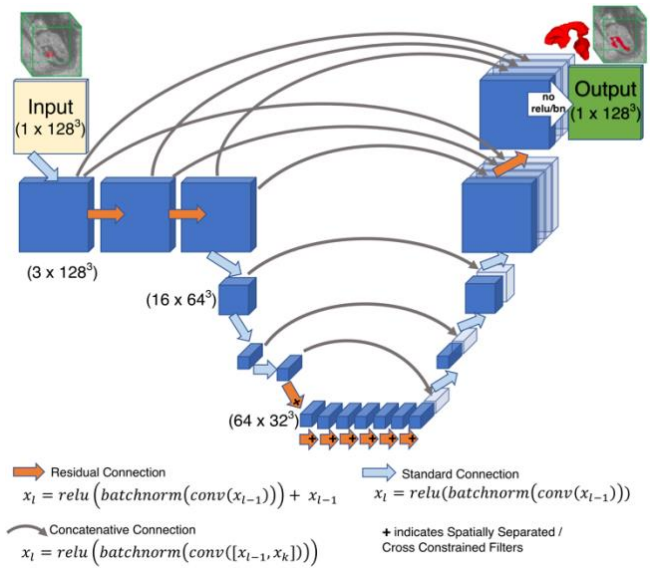

**Figure 4**: Pictorial representation of the segmentation network. $x_l$ = the $l_{th}$ layer activations. $x_0$ = input volume.

ReLU nonlinearities [16] and Batch Normalization [17] are applied after all convolutional layers except the final one to which a pointwise sigmoid nonlinearity is applied producing a probability of BV membership for each voxel.

### III. EXPERIMENTAL DESIGN AND ANALYSIS

#### DATASET

The dataset used in this work consists of 370 whole-body HFU images of mouse embryos in several stages of development (10-14.5 days in utero), which were obtained between the years of 2016 and 2018. All the volumetric ultrasound data were acquired in utero and in vivo from pregnant mice using a 5-element, 40-MHz annular array [2]. The dimensions of the 3D images range from 150×161×81 to 300×281×362 and the voxel size is 50×50×50 $\mu m$. For all the 370 images, manual segmentations of BVs were conducted by a trained research assistant using Amira [20] and verified by a small animal ultrasound imaging expert.

The 259 images obtained in 2017 are used as training data for both localization and segmentation, and the remaining 111 images obtained in 2016 and 2018 are used for testing.

**Table 1**: Localization results on 111 testing volumes.

| Methods | # detected boxes containing the entire BV | Proportion of BV voxels in the remaining detected boxes |
|---|---|---|
| Single classifier | 104 (93.7%) | 0.994, 0.981, 0.963, 0.968, 0.996, 0.817, 0.974 |
| Ensemble of 3 classifiers | 107 (96.4%) | 0.983, 0.990, 0.990, 0.784 |

#### LOCALIZATION AND SEGMENTATION SOFTWARE

For both localization and segmentation, we use PyTorch [21] to implement the deep neural networks. In this work, ITK-SNAP [22] is used to visualize some end-to-end segmentation results of our proposed framework.

#### TRAINING

We implement on the fly data augmentation with 90, 180, and 270 degree rotation around each axis as well as image flipping along each axis for the training of both the localization and segmentation networks.

#### LOCALIZATION TRAINING

To train the localization network, about half a million bounding boxes—divided approximately evenly between positive and negative examples—are extracted from the 259 training images. A weighted cross entropy loss with weight 1 for negative (not containing the BV) and 1.2 for positive is used to train the network to compensate for the imbalance between negative and positive samples. We train the network for 5 epochs using SGD with momentum 0.9 [23] and weight decay 1e-5. The initial learning rate is set to 0.01 and multiplied by 0.1 after the third epoch. Because of considerations of GPU memory and training efficiency, a batch size of 200 is used.

#### SEGMENTATION TRAINING

The Dice Similarity Coefficient measures the volumetric overlap between the model's predictions and the manual segmentations giving more weight to true positives [11]. The segmentation network is trained using a differentiable Dice-based loss function:

$$DSC = \frac{2\,\hat{Y} \cdot Y + \epsilon}{|\hat{Y}|_1 + |Y|_1 + \epsilon} \ .$$

$\hat{Y} \in [0,1]^{128 \times 128 \times 128}$ is tensor of order three in which each element indicates the probability predicted by the model that the corresponding voxel belongs to the BV. $Y \in \{0,1\}^{128 \times 128 \times 128}$ is a binary tensor of order three in which each element indicates the manual segmentation. $\epsilon$ is a small positive constant added for smoothness (set to 1e-4 in this work). $|\hat{Y}|_1$ indicates the sum over all elements in $\hat{Y}$ (in our case this corresponds to the expected number of voxels which belong to the BV).

From the original training set of 259 images all subvolumes of size 128×128×128 which contain at least 97% of the BV are extracted producing around 64k subvolumes for training. This roughly corresponds to the data augmentation of translation. The network is trained for 5 epochs each time sampling 22k subvolumes from the 64k extracted samples. The network is optimized using SGD with momentum 0.9 [23], weight decay 1e-5, and batch size 4. The initial learning rate is set to 0.01 and multiplied by 0.1 after the third epoch.

## IV. EXPERIMENTAL RESULTS AND DISCUSSION

### LOCALIZATION TESTING RESULTS

With a trained classifier, stride 3 is used to scan the down-sampled image. For each candidate window in each test image, if the probability assigned by the model is larger than 0.95 (tuned on the training set), it is classified as a positive example, which indicates that the current window contains the whole BV. Even with this high threshold there are typically many overlapping predicted bounding boxes. We further take the mean of the centroid positions of all detected BV boxes. The mean centroid coordinates are then up sampled to the original image scale and a 128×128×128 bounding box is cut around the detected centroid.

To further improve the classification accuracy, we train three networks with the same procedure (the parameters of the resulting networks converge to slightly different values in each instance due to random initialization, the use of SGD, and randomized data augmentation). We apply all three networks to each candidate box and use the arithmetic mean of their outputs as the predicted probability that the bounding box contains the BV. The classification results are summarized in Table 1. Even using the single classifier, the vast majority of the detected boxes contain 100% of the BV (104); the proportion contained in the remaining bounding boxes (7) is displayed in the rightmost column of Table 1.

### SEGMENTATION TESTING RESULTS

The trained segmentation network is then applied to the detected bounding box from the localization step. For each voxel, if the output probability of the network is larger than a threshold of 0.92 (tuned on the training set), it is classified as belonging to the BV.

**Table 2**: The mean DSC for the segmented BV by the NGC based framework [4] and our proposed method on the unseen 111 HFU volume test set.

| Methods | Mean DSC | # Failure Cases (DSC<0.6) |
|---|---|---|
| NGC-based [4] | 0.7119 | 21 |
| Deep BV (single model) | 0.8911 | 1 |
| Deep BV (ensemble) | 0.8951 | 0 |
| Deep BV (ensemble + post-processing) | **0.8956** | 0 |

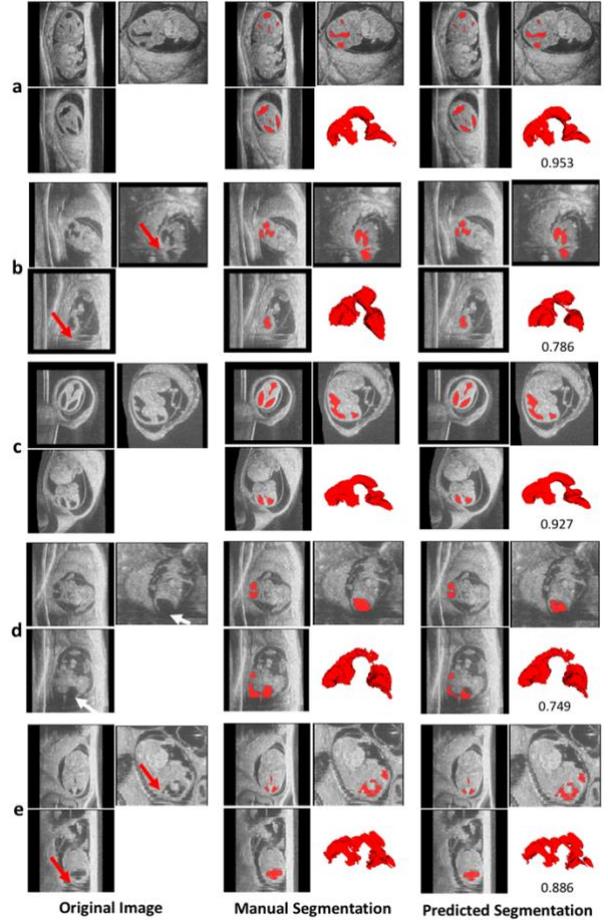

**Figure 5**: Visualization of 5 randomly selected examples. Each image is shown with coronal (top left), transverse (top right), sagittal (bottom left) and 3D BV (bottom right) views. The number below the BV 3D view is the corresponding Dice Similarity Coefficient. Note the missing boundaries in (d)-white arrows, and motion artifacts in (b,e)-red arrows (caused primarily by peristaltic motion of the mother's gastrointestinal system which perturbs the close by embryos).

To further improve the final results, we use a voxel-wise logical OR operation to combine the predictions of two independently trained segmentation networks. Because there are still some small isolated false positive regions in some images, we further apply connected component analysis post processing on the final output of the system and remove components with connected voxel number less than 300 (tuned on the training set).

As shown in Table 2, our final end-to-end results outperform the NGC-based framework [4] by a large margin both in terms of accuracy and robustness. As shown in Fig. 5, each of the images have different body orientation and BV location. Additionally, image (d) has a severe missing head boundary and images (b) and (e) have severe motion artifacts. However, our framework can still achieve competitive segmentation results when compared to manual segmentation, which clearly demonstrates the robustness of our method. We attribute this to the design of the segmentation architecture: (1) The fully convolutional design of the segmentation network allows end-to-end and pixel-to-pixel training. (2) The use of large kernels and a deep network stream enable a global receptive field allowing the network to consider global structure of the BV. These two factors allow the network to perform well even in the presence of missing head boundaries and severe motion artifacts.

V. SUMMARY

Segmentation of the BV in whole-body HFU images is a challenging task due to several factors including extreme class imbalance, which we mitigate by decomposing the task into the localization and segmentation steps, as well as the variety of body posture and orientation, BV shape, location and intensity, and presence of severe missing head boundaries and motion artifacts. Our proposed deep learning based method is fully automatic, robust to such conditions, and outperforms the previous state-of-the-art method by a margin of 25% in terms of mean DSC.

ACKNOWLEDGMENTS

The research described in this paper was supported in part by NIH grant EB022950.